\documentclass[pre,floatfix,twocolumn]{revtex4}  
\newcommand{\dri}{\delta r_i}
\usepackage{epsfig}
\usepackage{graphicx}

\begin{document}


\author{Brian Utter}
\author{R.P. Behringer}
\affiliation{Department of Physics and Center for Nonlinear
and Complex Systems, Box 90305,
Duke University, Durham, NC 27708}
\date{\today}
\title{Experimental Measures of Affine and Non-affine Deformation in Granular
Shear}

\renewcommand{\textfraction}{0.05}
\renewcommand{\topfraction}{0.95}
\renewcommand{\bottomfraction}{0.95}
\setcounter{bottomnumber}{4} 
\setcounter{topnumber}{4}
\renewcommand{\floatpagefraction}{0.95}

\newcommand{\ea}{{\it et al.}}

\begin{abstract}        
Through 2D granular Couette flow experiments, we probe failure and
deformation of disordered solids under shear.  Shear produces smooth
affine deformations in such a solid and also irresversible so-called
non-affine particle displacements.  We examine both processes.  We
show that the non-affine part is associated with diffusion, and also
can be used to define a granular temperature.  Distributions for
single particle non-affine displacements, $\dri$, satisfy $P_1(\dri)
\propto \exp [-|\dri/\Delta r|^{\alpha}]$ ($\alpha \stackrel{<}{\sim}
2$).  We suggest that the shear band forms due to a radially outward
diffusive flux/non-affine motion which is balanced in the steady state
by inward diffusion due to density gradients.

\end{abstract}


\maketitle


The way in which disordered solids undergo plastic i.e. irreversible
deformation is a topic of considerable recent
interest\cite{luding05,maloney04a,maloney04b,argon05,tanguy05,Falk:98:Dynamics,Lemaitre:02:Origin,Lemaitre:02:Rearrangements},
in part, because the phenomena appear to be general across a large
collection of materials that include glasses, foams, colloids and
granular materials, among others.  Here, we focus on granular
materials.  Despite a variety of experiments on bulk deformation of
granular materials, the microscopic basis for plasticity remains
elusive.  Molecular dynamics
simulations\cite{luding05,maloney04a,maloney04b,argon05,tanguy05} have
provided some insight, although a number of these studies involve
particles that interact via frictionless forces.  At this point, there
is relatively little experimental work that directly addresses the
connection between the microscopic and larger scale nature of granular
plasticity.  The goal of the present work is to begin to fill this
void.

\begin{figure}
\includegraphics[width=2.6in]{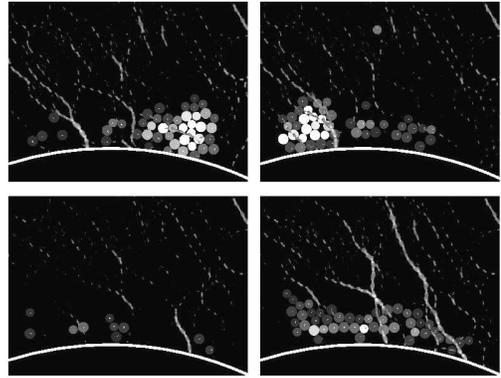}%
\caption{\label{fig:chains} Time sequence of images emphasizing the
  spatially inhomogeneous and temporally intermittent nature of force
  chains (bright lines) and active areas of non-affine deformation
  (shaded disks).  Each disk shows a local instantaneous value for
  $D^2_{min}$ (see text), where brighter corresponds to larger
  $D^2_{min}$.  Images are separated in time by $5\Delta t$ (see
  text), and run left-to-right, top-to-bottom.  The shearing wheel,
  located near the bottom of each image, rotates counterclockwise.
  Normally black, the edge of the wheel is highlighted for
  visability.}
\end{figure}

There is, however, a relatively extensive literature that considers
microscopic plastic deformation in amorphous molecular
solids\cite{maloney04a,maloney04b,argon05,tanguy05}. In the limit of
low temperatures where thermal noise is irrelevant, it is expected
that the plastic behavior in such materials is similar to what occurs
in a granular materials.  Both microscopic and granular materials
exist in disordered jammed states, and when sheared, undergo
irreversible configurational changes.  In the granular case, the
configurational structures also involve force chains, long filimentary
structures that form so as to resist shear,
e.g. Fig.~\ref{fig:chains}.

In recent theoretical work, Falk and Langer
(FL)\cite{Falk:98:Dynamics} developed models of elasto-plastic
deformation of amorphous solids which have also been applied recently
to granular materials
\cite{Lemaitre:02:Origin,Lemaitre:02:Rearrangements}.  FL consider
localized regions that are susceptible to irreversible
failure. Associated with any smooth affine structural change or
deformation in such a region, there is a non-affine part which leads
to irreversibility/plasticity. In their analysis, FL
\cite{Falk:98:Dynamics} consider small clusters of $N$ particles.
They fit local deformations occurring over a time $\Delta t$ to an
affine deformation.  To identify localized plastic rearrangements,
they define a quantity ($D^2_{min}$) which characterizes departures
from the local affine deformation.

In this letter, we characterize the motion generated in 2D granular
Couette (shear) flow in terms of affine and non-affine components.
For this system, deformation is localized to a shear band of width
less than 10 grain diameters.  Despite the relatively rapid variation
of quantities with radial distance, it is still possible to
characterize the local affine and non-affine components of the
flow. However, there are some differences, discussed below, between
the case of homogeneous pure shear studied in FL, and the present
case, which is spatially varying and also involves rotation of small
volume elements.

The experimental apparatus used here and described in greater detail
elsewhere\cite{Utter:04:Self}, is a 2D Couette shearing experiment.  A
dense packing of $\sim 45,000$ bidisperse photoelastic disks lie on a
flat horizontal surface.  In that plane, the disks are bounded by an
inner shearing wheel and an outer concentric ring.  The wheel rotates
slowly ($f =$~inverse rotation period~$ = 1 mHz$ in the present
experiments) to impose quasistatic shear.  Particles are marked with
lines for easier tracking of position and rotation.  The apparatus is
illuminated from below and imaged from above by two cameras.  One
camera observes the grains through crossed polarizers while the other
simultaneously takes unpolarized images.  In this way, we record both
stresses (through photoelastic measurement with polarized images) and
kinematic information (by particle tracking with unpolarized images).
We note that the motion is confined to a shear band near the shearing
wheel.  The tangential velocity $v(r)$ and local shear rate $\partial
v/\partial r$ decrease faster than exponentially with radial distance
from the shearing surface and have fallen by three orders of magnitude
from their value at the shearing wheel for $r \stackrel{>}{\sim}
8d$\cite{utter04b}.  Here, we choose $r = 0$ to correspond to the
surface of the shearing wheel.  In addition to the radial velocity
gradient, there is also a density gradient, i.e. Reynolds
dilatancy. The density/packing fraction near the wheel can be lower by
as much as 2/3 of the density far away.  

In order to characterize the motion, we consider localized regions
falling within $2.2 d$ of a reference particle, where $d$ is the mean
particle diameter.  (FL use a similar sampling radius of $2.5 a_{SL}$
where $a_{SL}$ is an interatomic distance.)  Typically, we find $N
\simeq 15$ particles in such a neighborhood, including the particle of
interest.  We then follow the motion of these $N$ particles over a
time $\Delta t = 1.85s$ that is short compared to $1/f$.  We are
interested in the deformation of this patch.  To this end, we
determine the location of each particle relative to the patch center
of mass (CoM), both initially ($r_i$), and after the small elapsed
time ($r_i'$).  We determine the smooth mapping that best describes
the affine deformation, by least squares fitting data for the $r_i$
and $r_i'$ in a cluster to the form $r_i' = {\bf E}r_i$, where ${\bf
E}$ is a $2 \times 2$ matrix.  $D^2_{min} = \Sigma (r_i' - {\bf
E}r_i)^2$ is the parameter of Falk and Langer, with the slight
modification that we must consider motion relative to the CoM, since
there is net flow in our system.  ${\bf E}$ is in general not
symmetric, and hence its eigenvalues may be (and often are) complex.
This is a consequence of the typical deformation in the present flows:
on average, there is both deformation of a volume element,
characterized by a symmetric tensor, ${\bf F}$, and rotation,
characterized by a rotation tensor, ${\bf R_{\theta}}$.  This differs
from the case of pure shear, where there is no rotation on average.
We write ${\bf E} = {R_{\theta}\bf F}$ without serious ambiguity.  For
instance, inverting the order of ${\bf F}$ and ${\bf R_{\theta}}$ does
not change $\theta$ or the eigenvalues of ${\bf F}$.  Also, $\theta$
is uniquely determined if we restrict $-\pi/2 < \theta \leq \pi/2$,
which is not an issue for the small deformations considered here.  We
can then express the deformation in terms of the eigenvalues,
$\epsilon_i$, of ${\bf \epsilon} = {\bf F - I} $, where ${\bf
\epsilon}$ is the conventional strain tensor, and ${\bf I}$ is the
unit tensor.  Together with $\theta$, they characterize the affine
part of the local evolution.  $D^2_{min}$ characterizes the extent of
the nonaffine motion for the small region in question.  We note that
all measures of the deformation, both affine and non-affine are
spatially and temporally intermittent, as seen for instance in
Fig.~\ref{fig:chains}.  Hence, it is important to consider the
probability distribution functions (PDF's) of these measures as
functions of $r$.

\begin{figure}
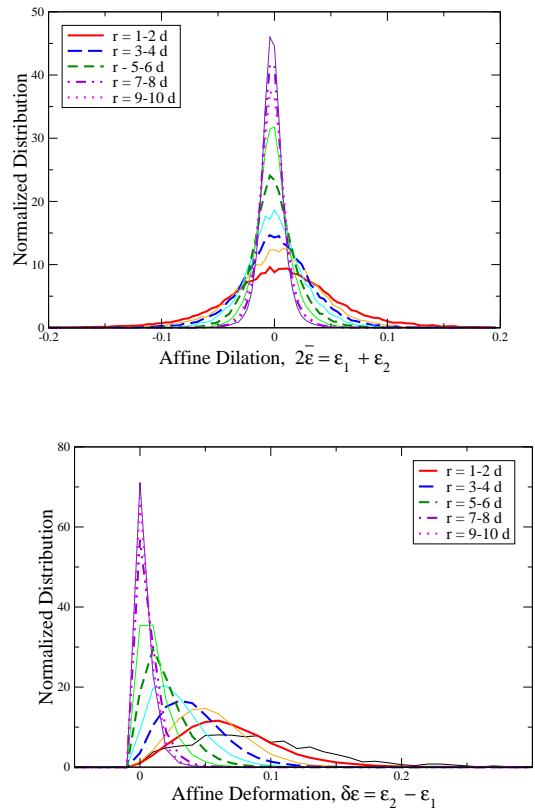


\includegraphics[width=2.6in]{./Fig2a.eps}%
\vspace{0.2in}
\newline

\includegraphics[width=2.6in]{./Fig2b.eps}%
\caption{\label{fig:affine} PDF's of the affine dilation, $\epsilon_1
  + \epsilon_2$ (top), and affine shear deformation, $\epsilon_2 -
  \epsilon_1$ (bottom).  Each series corresponds to a radial bin of
  width $d$.  The notation $r=1-2d$, etc. indicates that the central
  particle of the cluster was located radially between $d$ and $2d$
  from the wheel, etc. }
\end{figure}

We first consider data for the affine deformations.  In
Fig.~\ref{fig:affine} we show PDF's for the eigenvalues of ${\bf
\epsilon}$, which we give in the form $\delta \epsilon = \epsilon_2 -
\epsilon_1$ and $2 \bar{\epsilon} = \epsilon_1 + \epsilon_2$.  The
different data sets in each figure are binned over radial widths of
size $d$, and correspond to a series of radial distances, each $d$
apart, from the shearing wheel.  The first series is centered at $d/2$
from the shearing wheel.  We expect that $\bar{\epsilon}$ has zero
mean in the steady state, since it corresponds to the dilation of the
local patches; $\delta \epsilon$ has non-zero mean and a
characteristic shape that reappears in the data for $D^2_{min}$,
discussed below.  PDF's for $\theta$ resemble those for $\delta
\epsilon$; mean values for $\theta$ are typically a few degrees or
less.

\begin{figure}



\includegraphics[width=2.5in]{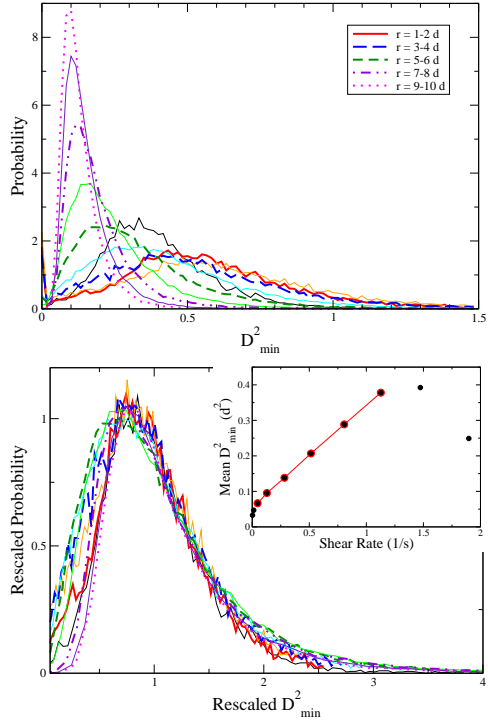}%
\newline

\caption{\label{STZdist} Top: PDF's of $D^2_{min}$ (units of $d^2$) at
various distances $r$ from the shearing surface (in units of $d$) for
1080 images.  Bottom: Rescaled data from top.  Each data set is fitted
and rescaled by the peak magnitude and mean value of $D^2_{min}$.
Inset: $\langle D^2_{min} \rangle$ vs. local shear rate.  Note the
drop in $\langle D^2_{min} \rangle$ (departure from red line) at high
$\dot{\gamma}$, i.e. near wheel. }
\end{figure}

\begin{figure}
\includegraphics[width=2.8in]{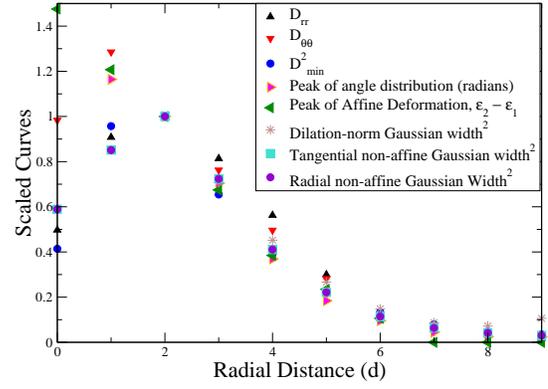}%
\caption{\label{STZvsdff} Rescaled $\langle D^2_{min} \rangle$, widths
  of non-affine PDF's, $P_1({\bf \dri})$, diffusivities, and peak
  values of $P(\theta)$ and $P(\epsilon_2 - \epsilon_1)$ versus radial
  distance from the shearing wheel.  All quantities are scaled to 1 at
  $r = 2d$.  For reference, at $r= 2d$, $\dot{\gamma} = 0.60s^{-1}$,
  $D_{rr} = 0.0054d^2/s$, $D_{\theta \theta} = 0.011d^2/s$,
  $<D^2_{min}> = 0.38d^2$, $\theta_{peak} = 0.045 rad$, $(\epsilon_2 -
  \epsilon_1)_{peak} = 0.047d$.  Square widths for other PDF's are:
  $P(\epsilon_2 -\epsilon_1)$--$0.0016d^2$, $(P_1(\delta
  r_i)--radial)^2$--$0.0050d^2$, $(P_1(\delta
  r_i)--azimuthal)$--$0.0053d^2$.}
\end{figure}

We plot the PDF's of $D^2_{min}$ binned at different distances $r$
from the shearing surface in Fig.~\ref{STZdist}(top).  The peak in the
PDF of $D^2_{min}$ tends to be at larger values of $D^2_{min}$ for $r$
closer to the shearing wheel, although the position of the peak does
not vary monotonically with $r$, a point that we consider below.  The
individual PDF's are well fitted by the form $f(x) = C_1 x^{C_2}
e^{-x/C_3}$, which we discuss further below. However, the exponent
$C_2$ and the scaling factor $C_3$ in the exponential vary from fit to
fit.  In Fig.~\ref{STZdist}(bottom), each of the data sets is rescaled
in the vertical direction by the peak fit value and in the horizontal
direction by the mean $\langle D^2_{min} \rangle$. We see that the
data at different $r$ collapse fairly well onto a single curve.  The
averages $\langle D^2_{min} \rangle$ (inset), are set by the local
shear rate within the shear band (inset, bottom).  We previously
observed a similar dependence for diffusivity, $D$, on local shear
rate $\dot{\gamma}$ , i.e. $D \propto \dot{\gamma}$
\cite{Utter:04:Self}.  In Fig.~\ref{STZvsdff}, we plot $\langle
D^2_{min} \rangle$ and radial and tangential diffusivities ($D_{rr}$
and $D_{\theta\theta}$) rescaled such that all quantities are $1$ at
$r=2d$.  The diffusivity is therefore proportional to the mean measure
of local plastic rearrangements $\langle D^2_{min} \rangle$.  It is
perhaps not surprising that diffusivity and $D^2_{min}$ are related,
as they both measure non-affine displacements.  Dimensionally,
$D^2_{min}$ has units of $x^2$ and is measured over a specified time
interval $\Delta t$.  Diffusivity $D$ is set by $D \sim \langle
(\Delta x)^2 \rangle/ \Delta t$.  In these experiments, displacements
scale by some fraction of $d$ over times scales of order the local
inverse shear rate $\dot{\gamma}^{-1}$, e.g. $D \sim \dot{\gamma}d^2$
and $D^2_{min} \sim \dot{\gamma} \Delta t d^2$.

The decrease in diffusivity and $\langle D^2_{min} \rangle$ for $r
\approx 0$ in Fig.~\ref{STZvsdff} results from several effects.
Particles in contact with the shearing surface are dragged
quasi-ballistically in the azimuthal direction\cite{Utter:04:Self}.
In addition, radial diffusion is limited to the outward direction next
to the shearing wheel.  Roughly, one might expect a decrease by 1/2
for particle variances for radial displacements (and hence radial
diffusivity, $D_{rr}$) near the wheel.

\begin{figure}
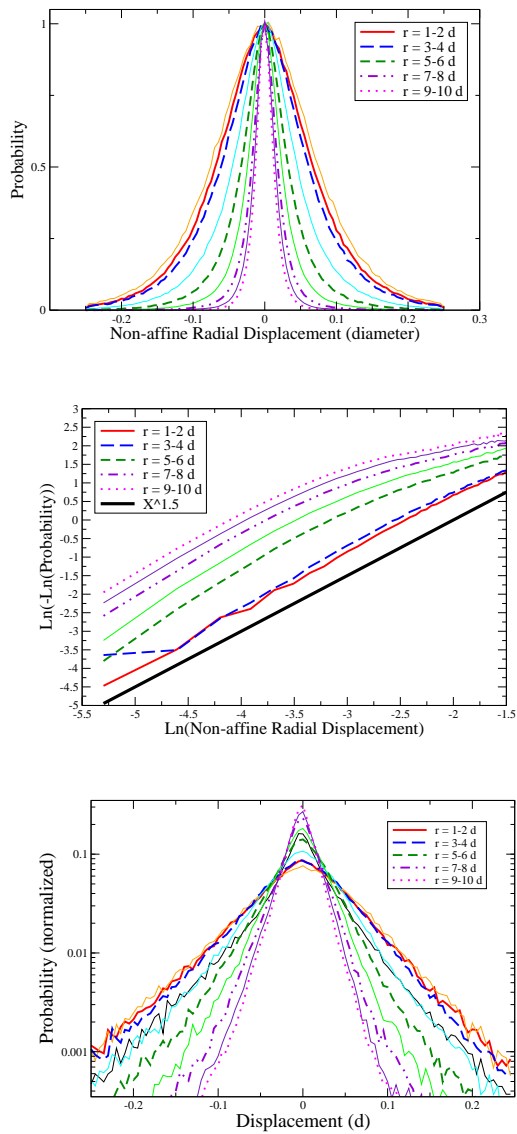

\includegraphics[width=2.5in]{./Fig5a.eps}%
\vspace{0.15in}
\newline

\includegraphics[width=2.5in]{./Fig5b.eps}%
\vspace{0.3in}
\includegraphics[width=2.5in]{./Fig5c.eps}%

\caption{\label{fig:p1} Data for individual particle non-affine radial
  displacements, $P_1(\dri)$.  Top: on linear scales, Middle: on
  loglog-log scales (solid black line shows slope 3/2).  PDF's for the
  non-affine azimuthal displacements are similar. Bottom: PDF's of
  total displacements, $r_i' - r_i$, on log-lin scales. }
\end{figure}

$D^2_{min}$ is a mesosopic property, i.e. a locally averaged measure
of a grain-scale process.  The scaling property observed for the PDF's
of $D^2_{min}$ suggest that there may be an underlying microscopic
phenomena of interest.  To this end, we examine in Fig~\ref{fig:p1}
(top), the PDF's $P_1({\bf \dri})$ of the individual non-affine
particle displacements, $\dri = r_i' -{\bf E}r_i$.  The PDF for $P_1$
is approximately gaussian, $P_1(\dri) \simeq A \exp [-(\dri/\Delta
r)^{\alpha}]$, where $\alpha \leq 2$.  To see this, we plot in
Fig.~\ref{fig:p1} (middle), $log |log (P_1 (\dri) )|$ as a function of
$log (\dri)$, with $P_1$ normalized so that $P_1(0) = 1$.  These data
are striking in as much as they resemble the velocity distributions
for granular gas-like states\cite{rouyer00,losert99}, even though the
basic kinetic theory assumption of short-lived uncorrelated
collisions/contacts is manifestly violated.  Interestingly, PDF's for
the {\em total} displacements, $r_i' -r_i$, are exponentially
distributed, Fig~\ref{fig:p1} (bottom).

$D^2_{min}$ is constructed by choosing $N$ particles in a cluster, and
then computing $D^2_{min} = (\Sigma_N \dri^2)$.  The PDF for
$D^2_{min}$, constructed from clusters of $N$ particles is
\begin{equation}
P(D^2_{min}) = \int P_N(\delta r_1, ..\delta r_N) \delta (D^2_{min} -\Sigma (\dri)^2) d\delta r_1...d\delta r_N,
\end{equation}
where $P_N$ gives the PDF for the $N$ non-affine particle
displacements.  We do not {\em a priori} know $P_N(\delta r_1,
..\delta r_N)$.  However, if we assume that the individual particle
displacements are uncorrelated and given by gaussian PDF's: $P_1(\dri)
= B\exp [-(\dri/\Delta r)^2]$, then the above integral for the PDF for
$D^2_{min}$ follows with a bit of algebra.  (Although neither of these
assumptions is rigorously true, neither is markedly unreasonable.)
Then, $P(D^2_{min}) = C_1 (D^2_{min})^{N-1} \exp (-D^2_{min}/C_2)$,
where the $C_i$ are constants.  This form is qualitatively consistent
with the experimentally obtained PDF's for
$D^2_{min}$ (Fig.~\ref{STZdist}).  Applying this strictly is probably
not warranted, since some correlation is likely, and the measured
$P_1$'s are not truly gaussians.

The widths of the $P_1(\dri)$ (Fig.~\ref{STZvsdff}) can be taken as an
effective granular temperature associated with the
non-affine/diffusive displacements of the particles (hence presumably
a configurational measure).  However, we give a note of caution here
concerning the relation of this temperature to the overall energy in
the system.  The elastic energy and elastic energy fluctuations for
this system are several orders of magnitude larger than the
corresponding kinetic energies associated with both the affine and
non-affine motion.  In addition, the motion considered here is driven
by the wheel displacement, not its velocity.  If the experiments were
slowed by any amount, the basic phenomena would appear essentially the
same when expressed in terms of displacements.

In closing we offer the following insight into the formation of a
shear band from an initially homogeneous state.  Both the affine and
non-affine motion are driven by and extract energy from the wheel
displacement and the stored elastic energy.  Non-affine deformation is
associated with diffusive motion, whose source is the stirring from
the wheel.  When the wheel begins to turn, there is a radially outward
flux driven by the excitation of the shearing wheel (there can be no
inward diffusion there).  This flux is the source of the (Reynolds)
dilation next to the wheel.  Such a flux is diffusive in the steady
state, and is then counterbalanced by the resulting gradients of
density/free volume which cause a compensating reverse mass flux.
This state is also mechanically stable due to the reduction in density
and contact number near the wheel.  The system is less resistant to
shear near the wheel due to contact number depletion, leading to
localization of motion.  In the present system, which has a fixed
volume, one expects an increase in density outside the dilated region,
which tends to limit the growth of the shear band, in contrast to
recent 3D experiments\cite{fenistein04} with an open top where the
growth of the shear band is much less constricted.  An interesting
question is the extent to which similar phenomena occur in other
disordered solids.

{\bf Acknowledgment} We appreciate helpful discussions with A.
Lemaitre and J. Langer, and comments from B. Chakraborty and M. Sperl.
This work has been supported by the NSF grants DMR-0555431,
DMS-0204677 and DMS-0244492, and NASA grant NAG3-2372.


\begin{thebibliography}{10}

\bibitem{Falk:98:Dynamics}
M. Falk and J. Langer, Physical Review E (Statistical Physics, Plasmas, Fluids,
  and Related Interdisciplinary Topics) {\bf 57},  7192  (1998).

\bibitem{luding05}
S. Luding, J. Phys.:Condens. Matter {\bf 17},  S2623  (2005).

\bibitem{maloney04a}
C. Maloney and A. Lema\^{i}tre, Phys. Rev. Lett. {\bf 93},  016001  (2004).

\bibitem{maloney04b}
C. Maloney and A. Lema\^{i}tre, Phys. Rev. Lett. {\bf 93},  195501  (2004).

\bibitem{argon05}
M.~J. Demkowicz and A.~S. Argon, Phys. Rev. B {\bf 72},  245206  (2005).

\bibitem{tanguy05}
F. Leonforte, R. Boissi\`{e}re, A. Tanguy, J.~P. Witmer, and J.-L. Barrat,
  Phys. Rev. B {\bf 72},  224206  (2005).

\bibitem{Lemaitre:02:Origin}
A. Lemaitre, Physical Review Letters {\bf 89},  064303  (2002).

\bibitem{Lemaitre:02:Rearrangements}
A. Lemaitre, Physical Review Letters {\bf 89},  195503  (2002).

\bibitem{Utter:04:Self}
B. Utter and R.~P. Behringer, Phys. Rev. E {\bf 69},  031308  (2004).

\bibitem{utter04b}
B. Utter and R.~P. Behringer, Euro. Phys. J. E {\bf 14},  373  (2004).

\bibitem{rouyer00}
F. Rouyer and N. Menon, Phys. Rev. Lett. {\bf 85},  3676  (2000).

\bibitem{losert99}
W. Losert, D.~G.~W. Cooper, J. Delour, A. Kudrolli, and J.~P. Gollub, Chaos
  {\bf 9},  682  (1999).

\bibitem{fenistein04}
D. Fenistein, J.~W. van~de Meent, and M. van Hecke, Phys. Rev. Lett. {\bf 92},
  094301  (2004).

\end{thebibliography}
\bibliographystyle{prsty}

\end{document}